\documentclass[11pt]{article}
\usepackage{amsmath,amssymb}
\newtheorem{theorem}{Theorem}
\begin{document}

\def\R{{\mathbb R}}
\def\C{{\mathbb C}}
\def\Z{{\mathbb Z}}
\def\Q{{\mathbb Q}}
\def\H{{\mathbb H}}
\def\D{{\cal D}}
\def\const{\mathrm{const}}

\title{Finite gap theory of the Clifford torus}
\author{Iskander A. TAIMANOV
\thanks{Institute of Mathematics, 630090 Novosibirsk, Russia;
e-mail: taimanov@math.nsc.ru}}
\date{}
\maketitle

\section{Introduction and main results}

In this paper we construct the spectral curve and the Baker--Akhiezer function
for the Dirac operator which form the data of the Weierstrass representation
of the Clifford torus. This torus appears in many conjectures 
from differential geometry (see Section \ref{sec2}).

By constructing this Baker--Akhiezer function we demonstrate 
a general procedure
for constructing Dirac operators and their
Baker--Akhiezer functions 
corresponding to singular spectral curves. This procedure is exposed in 
Section \ref{sec3}. 

The Clifford torus is a torus embedded into $\R^3$ which appears in many 
important problems of surface theory. The corresponding Dirac operator
is
$$
\D =
\left(
\begin{array}{cc}
0 & \partial \\
-\bar{\partial} & 0
\end{array}
\right)
+ 
\left(
\begin{array}{cc}
U & 0 \\
0 & U
\end{array}
\right)
$$
with the potential
\begin{equation}
\label{pot}
U(z,\bar{z}) = \frac{\sin y}{2\sqrt{2}(\sin y - \sqrt{2})}, \ \ \ 
z =x+iy.
\end{equation}

We have 

\begin{theorem}
\label{main}
The Baker--Akhiezer function of the Dirac operator $\D$ with the potential 
(\ref{pot}) is a vector function
$\psi(z,\bar{z},P)$, where $z \in \C$ and $P$ belongs to the spectral curve
$\Gamma$ of this operator, such that

1)
the spectral curve $\Gamma$ is a sphere $\C P^1 = \bar{\C}$ with 
two marked points $\infty_+ = (\lambda=\infty), \infty_- = (\lambda=0)$ 
where $\lambda$ is an affine parameter
on $\C \subset \C P^1$ and with two double points obtained by stacking together
the points from the following pairs:
$$
\left(\frac{1+i}{4},\frac{-1+i}{4}\right) 
\ \ \ \mbox{and} \ \ \ 
\left(-\frac{1+i}{4},\frac{1-i}{4}\right);
$$

2) 
the function $\psi$ is meromorphic on $\Gamma \setminus \{\infty_\pm\}$ and
has at the marked points (``infinities'') the following asymptotics:
$$
\psi \approx
\left(
\begin{array}{c}
e^{k_+  z} \\ 0
\end{array}
\right) \ \mbox{as $k_+ = \lambda \to \infty$};
\ \ 
\psi \approx
\left(
\begin{array}{c}
0 \\ e^{k_-  \bar{z}}
\end{array}
\right) \ \mbox{as $k_- = -\frac{|u|^2}{\lambda} \to \infty$}
$$
where $k^{-1}_\pm$ are local parameters near $\infty_\pm$ and 
$u=\frac{1+i}{4}$;

3)
$\psi$ has three poles on $\Gamma \setminus\{\infty_\pm\}$ which are 
independent on $z$ and have the form
$$
p_1 = \frac{-1+i + \sqrt{-2i-4}}{4\sqrt{2}}, \ \
p_2 = \frac{-1+i - \sqrt{-2i-4}}{4\sqrt{2}}, \ \
p_3 = \frac{1}{\sqrt{8}}.
$$

Therewith the geometric genus $p_g(\Gamma)$ and 
the arithmetic genus $p_a(\Gamma)$ of $\Gamma$ are as follows:
$$
p_g(\Gamma) = 0, \ \ \ \
p_a(\Gamma) = 2.
$$

The Baker--Akhiezer function satisfies
the Dirac equation 
$$
\D \psi = 0
$$
at any point of $\Gamma \setminus \{\infty_+,\infty_-,p_1,p_2,p_3\}$
where the potential $U$ of the Dirac operator takes the form (\ref{pot}).

The Clifford torus is constructed via the Weierstrass representation 
(\ref{weierstrass}) from the
function 
$$
\psi = \psi\left(z,\bar{z},\frac{1-i}{4}\right).
$$
\end{theorem}

Remark that in the proof of this theorem which will be given in Section 
\ref{sec4} it is actually will be showed that the Baker--Akhiezer function
$\psi$ takes the form
$$
\psi_1(z,\bar{z},\lambda) = e^{\lambda z -\frac{|u|^2}{\lambda}\bar{z}}
\left(
q_1 \frac{\lambda}{\lambda-p_1} + q_2 \frac{\lambda}{\lambda-p_2} +
(1-q_1-q_2)\frac{\lambda}{\lambda-p_3}\right),
$$
$$
\psi_2(z,\bar{z},\lambda) = e^{\lambda z -\frac{|u|^2}{\lambda}\bar{z}}
\left(
t_1 \frac{p_1}{p_1-\lambda} + t_2
\frac{p_2}{p_2-\lambda} +
(1-t_1-t_2)\frac{p_3}{p_3-\lambda}\right)
$$
where $u=\frac{1+i}{4}$ and $q_1,q_2,t_1,t_2$ are functions of $z,\bar{z}$
which are uniquely defined by the conditions
$$
\psi\left(z,\bar{z},\frac{1+i}{4}\right) = 
\psi\left(z,\bar{z},\frac{-1+i}{4}\right), \ \ \
\psi\left(z,\bar{z},-\frac{1+i}{4}\right) = 
\psi\left(z,\bar{z},\frac{1-i}{4}\right).
$$
It appears that $q_1,q_2,t_1,t_2$ 
are $2\pi$-periodic functions of $y$ and are independent on $x$.

We would like to mention one interesting feature:

\begin{itemize}
\item
the spectral curve $\Gamma$ admits a holomorphic involution 
$\sigma(\lambda)=-\lambda$ which preserves the infinities $\infty_\pm$.
Although both the Jacoby variety $J(\Gamma/\sigma)$ of the 
quotient space $\Gamma/\sigma$ and the Prym variety of this
involution are noncomplete Abelian varieties, the potential is a smooth
function.
\end{itemize}

This is explained by some effect unfamiliar for other operators.
It is as follows.

The potential $U$ is written in the terms of Prym functions
(as in the case of the two-dimensional Schr\"odinger operator and some other 
operators, see \cite{DKN2}). Although the Prym variety is isomorphic to
$\C^\ast$, the potential depends on one real-valued variable $y$ in a way 
that it is a restriction of some meromorphic function on the Prym
variety onto a compact circle $S^1 \subset \C^\ast$. This circle is a
compact subgroup of $\C^\ast$.

The correspondence between tori in $\R^3$ and Dirac operators $\D$ with
real-valued potentials is established by the Weierstrass representation.

It is based on a local representation of any surface immersed into $\R^3$
by the formulas
\begin{equation}
\label{weierstrass}
\begin{split}
x^k = x^k(0) + \int \left( x^k_z dz + \overline{x^k_z}
d\bar{z}\right), \ \ k=1,2,3, \\
x^1_z = \frac{i}{2}(\bar{\psi}^2_2 + \psi^2_1),
\ \ \
x^2_z = \frac{1}{2}(\bar{\psi}^2_2 - \psi^2_1),
\ \ \
x^3_z = \psi_1\bar{\psi}_2,
\end{split}
\end{equation}
where $\psi$ meets the Dirac equation
$$
\D \psi = 0.
$$
In this event $z$ defines a conformal parameter $z=x+iy$ on the surface,
the first fundamental form equals $e^{2\alpha} dz d\bar{z}$ and
$$
U = \frac{He^\alpha}{2}
$$
where $H$ is the mean curvature (see \cite{K2,T1}). 

After a globalization we obtain for a closed surface in
$\R^3$ a representation by these formulas where $\psi$ is a solution of the
Dirac equation and the Dirac operator acts on smooth sections of some
spinor bundles over a conformally equivalent surface of constant curvature
\cite{T1,T2}. 

It appears that the spectral curve
of $\D$ on the zero energy level defined initially for the two-dimensional
Schr\"odinger operator in \cite{DKN} has to have some geometric meaning.

Let us briefly recall the origin of the spectral curves in theory of 
differential operators with double-periodic coefficients.

The spectral curve $\Gamma$ is a complex curve which 
parameterizes the Floquet functions on the zero energy level, i.e.
joint eigenfunctions of $\D$ and translations $T_\gamma$ by periods
$$
T_\gamma f(z,\bar{z}) = f(z+\gamma,\bar{z}+\bar{\gamma}), 
\ \ U(z+\gamma,\bar{z}+\bar{\gamma})=U(z,\bar{z}).
$$
Here $D\psi=0$ (the ``eigenvalue'' equals zero) and $\psi$ is considered as a 
formal analytic solution to this equation 
not necessary belonging to some fixed functional space.
Floquet functions are glued together into a meromorphic function 
$\psi(z,\bar{z},P)$ on the
spectral curve. If $\Gamma$ is of finite genus, it is also completed
by two ``infinities'' at which $\psi$ has 
exponential asymptotics. This would be the Baker--Akhiezer function.
Given a pair of generators $\gamma_1,\gamma_2$ of the period lattice,
to every Floquet function $\psi(z,\bar{z},P)$ where corresponds two 
functions on the spectral curve holomorphic outside the ``infinities'', 
the multipliers
$\mu_1(P)$ and $\mu_2(P)$, $P\in \Gamma$, such that
$$
T_{\gamma_j}\psi(P) = \mu_j \psi(P).
$$

We correspond to a torus in $\R^3$ an operator $\D$ via 
the Weierstrass representation of the torus and the spectral curve of 
this operator and define
the spectral genus of a torus 
as the geometric genus of
the normalization of  the spectral curve (see \cite{T2}).

We conjectured that 

\begin{itemize}
\item
the spectral curve $\Gamma$ corresponding to a torus in $\R^3$
and the multipliers $\mu_1,\mu_2:\Gamma \to \C$ are 
invariant under conformal transformations of 
the ambient space $\R^3$ (that was confirmed in
\cite{GS});

\item 
given a conformal class of a torus, 
the Willmore functional attains its minima on tori with the minimal value of
the spectral genus.
\end{itemize}

In geometric problems the spectral curves related to integrable surfaces
are not always smooth. This was discussed for minimal tori in
$S^3$ in \cite{Hitchin}. For general tori in $\R^3$ we have to define
the spectral curves via Baker--Akhiezer functions $\psi$.
We show how this is done for the Clifford torus 
in Theorem \ref{main}.

The Willmore conjecture reads 
that the global minimum is attained on the Clifford torus for which 
the spectral genus vanishes (as it is showed by Theorem \ref{main}).
We think that

\begin{itemize}
\item 
given a conformal class of a torus, 
the Willmore functional attains its minima on tori with the minimal value of
the spectral genus and the minimal value of the arithmetic genus of 
the spectral curve $\Gamma_\psi$ defined via the Baker--Akhiezer function.
\end{itemize}

We shall discuss this conjecture elsewhere together with the
expected relation between the spectral curves of tori in $S^3$ (as they are
defined in \cite{T4}) and the spectral curves of their stereographic images
in $\R^3$. For the Clifford torus
the mapping
$$
\C^\ast \to \C^\ast /\left\{\frac{1+i}{4} \sim \frac{-1+i}{4}, 
-\frac{1+i}{4} \sim \frac{1-i}{4}\right\}
$$
establishes an isomorphism of the corresponding 
spectral curves and the equivalence of 
the multipliers $\mu_1,\mu_2$. This was the basic idea of the computations
in Section \ref{sec4}, i.e. in the proof of Theorem \ref{main}.

We summarize the contents of the paper:

in Section \ref{sec2} we explain the geometry of Clifford torus in $\R^3$;

in Section \ref{sec3} we expose the procedure for constructiing Dirac operators
and their Baker--Akhiezer functions corresponding to singular spectral curves;

in Section \ref{sec4} we prove Theorem \ref{main}.

This work was supported by RFBR (grant 03-01-00403),
the Programme ``Leading Scientific Schools'' (NS-2185.2003.1), and
Max-Planck-Institute on Mathematics in Bonn.

\section{The Clifford torus}
\label{sec2}

In fact, in differential geometry two objects are called the Clifford torus:

1) (a torus in $S^3$) the product of circles of the same radii which lies 
in the unit $3$-sphere $S^3 \subset \R^4$ and therewith is defined by the
equations:
$$
x_1^2 + x_2^2 = x_3^2 + x_4^2 = \frac{1}{2}
$$
where $(x_1,\dots,x_4)$ are Euclidean coordinates in $\R^4$;

2) (a torus in $\R^3$) 
the following torus of revolution: take in the $x_1 x_3$ 
plane a circle $\gamma$
of radius $r = 1$ such that the distance between its center and the $x_1$ axis 
equals $R=\sqrt{2}$ and obtain a torus of revolution in $\R^3$ 
by rotating this circle 
$\gamma$ around the $x_1$ axis.

The torus in $S^3$ is considered up to isometries of $S^3$ and in symplectic 
geometry it is considered as a torus in a symplectic $4$-space $\R^4$ 
and is also called by this name.  

The torus in $\R^3$ is considered up to conformal transformations of $\R^3$.
In this event it is distinguished among tori of revolution by the ratio
$R/r = \sqrt{2}$.

These tori are related to many conjectures in geometry:

\begin{itemize}
\item
the Willmore functional defined on immersed surfaces in $\R^3$ by the formula
$$
{\cal W}(M) = \int_M (H^2-K) d\mu,
$$
where $d\mu$ is the area form in the induced metric, $H$ and $K$ are the mean 
curvature and the Gaussian curvature, respectively, for tori is not less than
$2\pi^2$ and attains its minimum on the Clifford torus (and its images under
conformal transformations of $\R^3$) (the Willmore conjecture);

\item
the Clifford torus in $S^3$ is the unique (up to isometries)
embedded minimal torus in $S^3$ (the Lawson conjecture);

\item
the area of any minimal torus in $S^3$ is not less than $2\pi^2$ 
which is the area of the Clifford torus;

\item
the Clifford torus in $S^3 \subset \R^4$ minimizes the Willmore functional 
$$
{\cal W}(M) = \frac{1}{4} \int_M |{\bf H}|^2 d\mu,
$$ 
where ${\bf H}$ is the mean curvature vector, for tori in $\R^4$, or at least
for Lagrangian tori in $\R^4$ with the standard symplectic structure;

\item
the Clifford torus in $S^3 \subset \R^4$ minimizes the area in its Hamiltonian
isotopy class of Lagrangian tori in $\R^4$,
i.e. among tori obtained from the Clifford torus by Hamiltonian deformations
(the Oh conjecture).
\end{itemize}

The relation between two different notions of the Clifford torus 
and the corresponding conjectures is supplied by the stereographic projection
$F$ of $S^3$ onto $\R^3 \cup \{\infty\} = \bar{\R}^3$. 
For that project $S^3
\setminus \{(0,0,0,1)\}$ onto $\R^3 = \{x_4=0\}$ from the north pole 
$N=(0,0,0,1) \in S^3$ mapping $N$ into $\infty \in \bar{\R}^3$.
This mapping establishes the conformal equivalence of $S^3$ and $\bar{\R}^3$
and in coordinates takes the form
$$
F(x_1,x_2,x_3,x_4) = \left(\frac{x_1}{1-x_4},\frac{x_2}{1-x_4},
\frac{x_3}{1-x_4}\right), \ \ \ x_1^2+x_2^2+x_3^2+x_4^2 = 1.
$$
It is known that for a minimal torus $T^2 \subset S^3$ its area coincides with
the value of the Willmore functional on its stereographic image:
$$
\mathrm{Area}(T^2) = {\cal W}(F(T^2)).
$$
Therefore the third conjecture follows from the Willmore conjecture
although The opposite is not true (these conjectures are not equivalent).
Moreover the Lawson conjecture also implies the third conjecture 
(for expositions of the conjectures, some related results and their relations 
for tori in $\R^3$ see, for instance, \cite{B,T1} and 
for tori in $\R^4$ see \cite{M}).

Let us parameterize the Clifford torus in $S^3$ as follows:
$$
x_1 = \frac{\cos x}{\sqrt{2}}, \ x_2 = \frac{\sin x}{\sqrt{2}}, \ 
x_3 = \frac{\cos y}{\sqrt{2}}, \ x_4 = \frac{\sin y}{\sqrt{2}}
$$
where $0 \leq, x,y \leq 2\pi$.
Then the Clifford torus in $\R^3$ is given by the formulas
\begin{equation}
\label{clifform1}
r(x,y) = \left(\frac{\cos x}{\sqrt{2} - \sin y},
\frac{\sin x}{\sqrt{2} - \sin y},
\frac{\cos y}{\sqrt{2} - \sin y}\right).
\end{equation}
It is easy to compute that $x,y$ define a conformal parameter $z=x+iy$ on
the torus in which the induced metric takes the form
$$
ds^2 = \frac{dz d\bar{z}}{(\sqrt{2}-\sin y)^2},
$$
the normal vector (assuming that the orientation on the surface is defined
by the positively oriented frame $(r_x,r_y)$) equals
$$
N = \left(
\frac{\cos x (1-\sqrt{2}\sin y)}{\sqrt{2}-\sin y},
\frac{\sin x (1-\sqrt{2}\sin y)}{\sqrt{2}-\sin y},
\frac{-\cos y}{\sqrt{2}-\sin y}
\right),
$$
the second fundamental form is
$$
\frac{(\sqrt{2}\sin y - 1)}{(\sqrt{2} - \sin y)^2} dx^2 + 
\frac{1}{(\sqrt{2} - \sin y)^2} dy^2,
$$
the principal curvatures take very simple forms:
$$
\kappa_1 = 1, \ \ \kappa_2 = \sqrt{2} \sin y - 1,
$$
and therefore the mean curvature equals
$$
H = \frac{\sin y}{\sqrt{2}}.
$$

The potential of the Weierstrass representation is
\begin{equation}
\label{clifpotential}
U = \frac{\sin y}{2\sqrt{2}(\sqrt{2}-\sin y)}.
\end{equation}
Other important functions related to this representation are
$$
\frac{\partial}{\partial z} ( x_1+ix_2) = i\bar{\psi}_2^2, \ \ \
\frac{\partial x_3}{\partial z} = \psi_1 \bar{\psi}_2
$$
and for the Clifford torus they are equal to
\begin{equation}
\label{clifform2}
\begin{split}
\frac{\partial}{\partial z} ( x_1+ix_2) = 
\frac{\partial}{\partial z}
\frac{\cos x+ i \sin x}{\sqrt{2}-\sin y} = 
\frac{i}{2} e^{ix} \frac{\sqrt{2}-\cos y - \sin y}{(\sqrt{2}- \sin y)^2},
\\
\frac{\partial x_3}{\partial z} = 
\frac{\partial}{\partial z}
\frac{\cos y}{\sqrt{2}-\sin y} =
\frac{i}{2} \frac{\sqrt{2} \sin y -1}{(\sqrt{2} - \sin y)^2}.
\end{split}
\end{equation}

\section{Baker--Akhiezer functions and Dirac operators}
\label{sec3}

Let us recall the definition of the Baker--Akhiezer (vector) function $\psi$
corresponding to the Dirac operator
$$
{\cal D}
=
\left(
\begin{array}{cc}
0 & \partial \\
-\bar{\partial} & 0
\end{array}
\right)
+
\left(
\begin{array}{cc}
U & 0 \\
0 & V
\end{array}
\right).
$$
By definition it depends on a complex variable $z \in \C$ and on a parameter
$P$ on a complex curve $\Gamma$ of finite arithmetic genus
$g=p_a(\Gamma)$ and meets the following conditions:

1) $\psi$ is meromorphic in $P$ outside a couple of marked points
$\infty_\pm \in \Gamma$ and has poles at $g+1$ points
$P_1+ \dots + P_{g+1}$ (the points $\infty_+,\infty_-,
P_1,\dots,P_{g+1}$ are nonsingular);

2) $\psi$ has the following asymptotics near $\infty_\pm$:
$$
\psi \approx
e^{k_+ z}
\left[
\left(
\begin{array}{c}
1 \\ 0
\end{array}
\right)
+
\left(
\begin{array}{c}
\xi^+_1 \\ \xi^+_2
\end{array}
\right)
k_+^{-1}
+
O(k_+^{-2})
\right]
\ \ \mbox{as $P \to \infty_+$},
$$
$$
\psi \approx
e^{k_- \bar{z}}
\left[
\left(
\begin{array}{c}
0 \\ 1
\end{array}
\right)
+
\left(
\begin{array}{c}
\xi^-_1 \\ \xi^-_2
\end{array}
\right)
k_-^{-1}
+
O(k_-^{-2})
\right]
\ \ \mbox{as $P \to \infty_-$},
$$
where $k^{-1}_\pm$ are local parameters near infinities such that
$$
k^{-1}_\pm(\infty_\pm) = 0.
$$
It follows from the general theory of Baker--Akhiezer functions
(\cite{Krichever,DKN2}) that
the such function is unique (for a generic divisor
$D = P_1 + \dots + P_{g+1}$) and therefore we can find
an operator $\D$ with potentials $U$ and $V$ such that $\D \psi$
is meromorphic with analogous singularities but with the asymptotics
$$
\D \psi = e^{k_+ z} O(k_+^{-1}) \ \ \mbox{as $P \to \infty_+$}, \ \ 
\D \psi = e^{k_- \bar{z}} O(k_-^{-1}) \ \ \mbox{as $P \to \infty_-$}.
$$
The uniqueness of $\psi$ implies that $\D \psi=0$.
We have

\begin{theorem}[\cite{T2,T3}]
\label{theorem1}
The Baker--Akhiezer function $\psi$ satisfies the Dirac equation
$$
\D \psi = 0 \ \ \ \mbox{with}\ \ \ U = -\xi^+_2, \ \ V =\xi^-_1.
$$
\end{theorem}

If the curve $\Gamma$ is singular
we assume that it has some special form.
Let us now define what it is this form and recall some algebro-geometric
properties of such singular curves following \cite{Serre}.

Let $\Gamma_{\mathrm{nm}}$ be a nonsingular complex curve.
Take on $\Gamma_{\mathrm{nm}}$ effective divisors $D_1,\dots,D_n$, i.e.
for any $k=1,\dots,k$ such a divisor $D_k$ is
a formal sum of finitely many points on $\Gamma_{\mathrm{nm}}$ with positive
coefficients:
$$
D_k = a_{k1} Q_{k1} + \dots + a_{km_k} Q_{km_k},
$$
$$
Q_{kj} \in \Gamma_{\mathrm{nm}}, a_{kj} > 0, a_{kj} \in \Z,
j=1,\dots,m_k,k=1,\dots,n.
$$
We assume that the supports $\mathrm{supp}D_k = Q_{k1} + \dots + Q_{km_k}$,
$k=1,\dots,n$, of these divisors
are pairwise nonintersecting. Then we denote by
$$
\Gamma_{D_1,\dots,D_n}
$$
the singular curve obtained by contracting  all points from each divisor
$D_k$ into one singular point $S_k \in \Gamma$. This is done with respect
to the multiplicities. This means that
a function $f$ which is meromorphic on $\Gamma$
is, by definition, a meromorphic function
$f: \Gamma_{\mathrm{nm}} \to \C$
such that it has no poles at $\cup_k \mathrm{supp} D_k$,
and for each divisor
$$
D_k = a_{k1} Q_{k1} + \dots + a_{km_k} Q_{km_k}
$$
we have
$$
f(Q_{k1}) = \dots = f(Q_{km_k}), \ \
\partial^m f(Q_{kl}) = 0 \ \ \mbox{for $l=1,\dots,(a_{kl}-1)$},
$$
where $k=1,\dots.n$. Here $\partial$ is the Cauchy derivation with respect
to a complex parameter on $\Gamma$.

The natural projection
$$
\pi: \Gamma_{\mathrm{nm}} \to \Gamma
$$
is the normalization mapping.

If $\Gamma$ is singular we assume that it takes the form
$$
\Gamma = \Gamma_{D_1,\dots,D_n}.
$$

Let $T \subset \Gamma_{D_1,\dots,D_n}$ and $T$ is empty or contains only
nonsingular points.
Recall that a one-form $\omega$ is called a regular form
on $\Gamma_{D_1,\dots,D_n} \setminus T$
if $\omega$ is a meromorphic form on
$\Gamma_{\mathrm{nm}} \setminus \pi^{-1}(T)$ which may have poles
only at the supports of $D_1,\dots,D_k$  and at each point $Q_{kl}$
the degree of the pole is not greater than
$a_{kl}$ and moreover for every $k$ the inequality
$$
\sum_l \mathrm{Res} f\omega (Q_{kl}) = 0
$$
which holds for all meromorphic functions on $\Gamma$.
For a nonsingular curve
$\Gamma$ the notions of regular and holomorphic forms coincide.

For nonsingular curves the values of the arithmetic genus $p_a$ and
the geometric genus $p_g$ coincide. For a singular curve $\Gamma =
\Gamma_{D_1,\dots,D_k}$ we have
$$
p_g(\Gamma) = p_g(\Gamma_{\mathrm{nm}}), \ \
p_a(\Gamma) = p_g(\Gamma) + \sum_k (\deg D_k -1)
$$
where the degree of the divisor $D_k$ equals to
$$
\deg D_k = \deg (a_{k1} Q_{k1} + \dots + a_{km_k} Q_{km_k}) =
a_{k1} + \dots + a_{km_k}.
$$

We say that $\sigma: \Gamma \to \Gamma$ is an involution of $\Gamma$
if $\sigma^1=1$ and its pull-back defines an
involution on $\Gamma_{\mathrm{nm}}$
such that the singularity divisors $D_1,\dots,D_n$
falls into two groups:

1) divisors which are preserved: $D_j=\sigma(D_j)$;

2) divisors which are interchanges with others: $D_j = \sigma(D_k), j\neq k$.

The first group corresponds to fixed points $S_j=\pi(D_j)$
of an involution on $\Gamma$
although the points from $D_j$ could be permuted.

\begin{theorem}[\cite{T2,T3}]
\label{theorem2}
1) Let $\sigma$ be a holomorphic involution
$\sigma: \Gamma \to \Gamma$
such that
$$
\sigma(\infty_\pm) = \infty_\pm, \ \ \ \
\sigma(k_\pm) = -k_\pm
$$
under which the singular points $S_1=\pi(D_1),\dots,S_k=\pi(D_k)$
are fixed and other singular points are no fixed.

If there exists a meromorphic differential $\omega$ on
$\Gamma_{2D_1,\dots,2D_k,D_{k+1},\dots,D_n}$ such that
$\omega$ has two poles at
$\infty_\pm$ with the principal parts
$\pm k_\pm^2 (1 + O(k^{-1}_\pm)) dk^{-1}_\pm$,
it is regular outside $\infty_\pm$ and it has zeros at $D + \sigma(D)$
then the potentials $U$ and $V$ coincide: $U=V$.

2) Let $\tau$ be an antiholomorphic involution
$\tau: \Gamma \to \Gamma$
such that
$$
\tau(\infty_\pm) = \infty_\mp, \ \ \ \
\tau(k_\pm) = -\bar{k}_\mp
$$
under which the singular points $S^\prime_1=\pi(D^\prime_1),\dots,
S^\prime_l=\pi(D^\prime_l)$
are fixed and other singular points are no fixed
(here $(D^\prime_1,\dots,D^\prime_k)$ is obtained by a permutation of the
set $(D_1,\dots,D_n)$).

If there exists
a meromorphic differential $\omega^\prime$ on $\Gamma_{2D^\prime_1,\dots,
2D^\prime_l,D^\prime_{l+1},\dots,D^\prime_n}$ such that
it has two poles at
$\infty_\pm$ with the principal parts
$k_\pm^2 (1 + O(k^{-1}_\pm)) dk^{-1}_\pm$,
it is regular outside $\infty_\pm$ and it has zeros at $D + \tau(D)$,
then the potentials $U$ and $V$ are real-valued: $U=\bar{U}, \ V = \bar{V}$.
\end{theorem}

{\sc Proof.} We proved this theorem for smooth curves in \cite{T2} and for
singular curves for which all singular points are fixed by involution in
\cite{T3} (here we correct some inaccuracy and typos in the conditions on
the antiholomorphic involution). In general, the proof is the same for
all these cases:
we have to compute the residues of the following differentials
with poles only in $\infty_+$ and $\infty_-$
and remember that their sum taken over all poles vanishes:

1) the sum of the residues of the differential
$\psi_1(P)\psi_2(\sigma(P))\omega$
equals $-2\pi i(\xi^+_2+\xi^-_1) =0$ which implies that $U=V$;

2) the sums of the residues of the differentials
$\psi_1(P)\overline{\psi(\tau(P))}\omega^\prime$ and
$\psi_2\overline{\psi_2(\tau(P))}\omega^\prime$ are equal to
$2\pi i(\xi^-_1-\bar{\xi}^-_1)$ and
$2\pi i(\xi^+_2-\bar{\xi}^+_2$ respectively
which implies that $U = \bar{U}$ and $V=\bar{V}$.

This proves the theorem.

There is a general procedure for constructing Baker--Akhiezer functions
corresponding to different operators and smooth curves $\Gamma$
\cite{Krichever}. In \cite{T2} we applied it to the Dirac operator and
derived the explicit formulas for $\psi$ and the potentials $U$ and $V$ in
terms of the theta function of $\Gamma$.

For $\Gamma=\Gamma_{D_1,\dots,D_n}$ we have to do the following:

\begin{itemize}
\item
take a divisor $D=P_1+\dots+P_s$ of degree $s = p_a(\Gamma)+1 =
p_g(\Gamma_{\mathrm{nm}}) + \sum_k (\deg D_k - 1 ) + 1$;

\item
for each divisor $D^\prime_j= P_1+\dots+P_g+P_{g+j}$
where $g=p_g(\Gamma) = p_a(\Gamma_{\mathrm{nm}})$ and $j=1,\dots,
r=\sum_k(\deg D_k -1)$ construct the Baker-Akhiezer $\varphi_j$
on $\Gamma_{\mathrm{nm}}$ with $D^\prime_j$ as the divisor of poles
(as it was done in \cite{T2});

\item
construct the function $\psi$ in the form
$$
\left(
\begin{array}{c}
\psi_1 \\ \psi_2
\end{array}
\right)
=
\left(
\begin{array}{c}
q_1 \varphi_{1,1} + \dots + q_r \varphi_{r,1} \\
t_1 \varphi_{1,2} + \dots + t_r \varphi_{r,2}
\end{array}
\right)
$$
by solving the following equations on $q_i,t_j$:
$$
\psi_\alpha(Q_{k1}) = \dots = \psi_\alpha(Q_{km_k}),  \ \ \
\partial^m \psi_j(Q_{kl}) = 0, \ \ \
$$
$$
\sum_i q_i = \sum_j t_j = 1
$$
where $k=1,\dots,n$, $\alpha=1,2$,
$l=1,\dots,m_k$, and $m=1,\dots,(a_{kl}-1)$.
\end{itemize}

Notice that we have exactly $2r$ equations on the same number parameters.
The first two systems means that $\psi$ is defined on
$\Gamma$ and the latter two equations
implies the correct asymptotics of $\psi$ at the infinities.
The solutions $q_i,t_j$ depend on the spatial parameter $z$
as well as on the parameter on $\Gamma$.

Remark that complex curves with more general singularities
were considered in \cite{DKMM} as the spectral curves of the operator
$L = i\partial_y - \partial^2_x + u$. 
However we do not know do such singularities appear on the spectral curve of
a Dirac operator and if they do what have to be analogs of the conditions on
the involutions $\sigma$ and $\tau$. In our case such conditions involve 
not only the spectral curve $\Gamma_{D_1,\dots,D_n}$ but also another curve
$\Gamma_{2D_1,\dots,2D_k,D_{k+1},\dots,D_n}$.

\section{The spectral curve of the Clifford torus}
\label{sec4}

In this section we prove Theorem \ref{main}.

Let $\psi(z,\bar{z},\lambda)$ be a function on $\C P^1$
which is meromorphic in $\lambda \in \C \subset \C P^1$ on
$\C P \setminus\{0,\infty\}$, has a pole at the point $\lambda=p$ and
has the following asymptotics at two ``infinities'':
$$
\psi \approx
\left(
\begin{array}{c}
e^{k_+  z} \\ 0
\end{array}
\right) \ \ \mbox{as $k_+ = \lambda \to \infty$};
\ \ \
\psi \approx
\left(
\begin{array}{c}
0 \\ e^{k_-  \bar{z}}
\end{array}
\right) \ \ \mbox{as $k_- = -\frac{|u|^2}{\lambda} \to \infty$}.
$$
This means that it is a Baker--Akhiezer (vector)
function corresponding to the data
$$
\Gamma_u = \C P^1, D=p, \infty_\pm, k_\pm.
$$
It is known from the general theory that such a function is unique and
we easily compute it obtaining the following formula:
$$
\psi =
\frac{\lambda}{\lambda-p}\,
e^{\lambda z - \frac{|u|^2}{\lambda}\bar{z}}
\left(
\begin{array}{c}
1 \\ -\frac{p}{\lambda}
\end{array}
\right).
$$
The function $\psi$ satisfies the Dirac equation
$$
\D \psi = 0
$$
where the potentials of the Dirac operator $\D$ are
$$
U = p, \ \ \ V = \frac{|u|^2}{p}.
$$
If $p=u$ we have the operator
$$
\D =
\left(
\begin{array}{cc}
0 & \partial \\ -\bar{\partial} & 0
\end{array}
\right)
+
\left(
\begin{array}{cc}
u & 0 \\ 0 & \bar{u}
\end{array}
\right).
$$

We have two involutions on $\Gamma_u$: a holomorphic involution $\sigma$ and
an antiholomorphic involution $\tau$ which act as follows
$$
\sigma(\lambda) = \lambda, \ \ \
\tau(\lambda) = \frac{|u|^2}{\bar{\lambda}}.
$$

We shall look for the spectral data of the Clifford torus
in the following form:

\begin{itemize}
\item
the normalized spectral curve $\Gamma_{\mathrm{nm}}$ is $\C P^1$
with two infinities $\infty_\pm$;

\item
the actions of $\sigma$ and $\tau$ descend to involutions on $\Gamma$;

\item
$u = \frac{1+i}{4}$.
\end{itemize}

Consider the following singular curve:
$$
\Gamma = \C P^1/\{\pm u \sim \mp\bar{u}\}, \ \ \ u=\frac{1+i}{4},
$$
i.e. it is obtained from $\C P^1$ by gluing $u$ with $-\bar{u}$ and
$-u$ with $\bar{u}$. We see that $\Gamma$
is a sphere with a pair of double points.
We have
$$
p_g(\Gamma) = 0, \ \ p_a(\Gamma)=2.
$$
Therefore we look for the Baker--Akhiezer function on $\Gamma$ in the form
$$
\psi_1(z,\bar{z},\lambda) = e^{\lambda z -\frac{|u|^2}{\lambda}\bar{z}}
\left(
q_1 \frac{\lambda}{\lambda-p_1} + q_2 \frac{\lambda}{\lambda-p_2} +
(1-q_1-q_2)\frac{\lambda}{\lambda-p_3}\right),
$$
$$
\psi_2(z,\bar{z},\lambda) = e^{\lambda z -\frac{|u|^2}{\lambda}\bar{z}}
\left(
t_1 \frac{p_1}{p_1-\lambda} + t_2
\frac{p_2}{p_2-\lambda} +
(1-t_1-t_2)\frac{p_3}{p_3-\lambda}\right)
$$
where $p_1,p_2,p_3$ are the poles of $\psi$ and $q_1,q_2,t_1$, and
$t_2$ are functions of $z,\bar{z}$ obtained from the conditions:
$$
\psi(z,\bar{z},u) = \psi(z,\bar{z},-\bar{u}), \ \ \
\psi(z,\bar{z},-u) = \psi(z,\bar{z},\bar{u}).
$$
This leads to the following equations for $q_1$ and $q_2$:
$$
q_1
\left[
e^{uz-\bar{u}\bar{z}} u
\left(\frac{1}{u-p_1}-\frac{1}{u-p_3}\right)
-
e^{u\bar{z}-\bar{u}z} \bar{u}
\left(\frac{1}{\bar{u}+p_1} - \frac{1}{\bar{u}+p_3}\right)
\right] +
$$
$$
+
q_2
\left[
e^{uz-\bar{u}\bar{z}} u
\left(\frac{1}{u-p_2}-\frac{1}{u-p_3}\right)
-
e^{u\bar{z}-\bar{u}z} \bar{u}
\left(\frac{1}{\bar{u}+p_2} - \frac{1}{\bar{u}+p_3}\right)
\right] =
$$
$$
=
- e^{uz-\bar{u}\bar{z}} \frac{u}{u-p_3}
+
e^{u\bar{z}-\bar{u}z} \frac{\bar{u}}{\bar{u}+p_3},
$$
$$
q_1
\left[
e^{\bar{u}z - u\bar{z}} \bar{u}
\left(\frac{1}{\bar{u}-p_1} - \frac{1}{\bar{u}-p_3}\right)
-
e^{\bar{u}\bar{z}-uz} u
\left(\frac{1}{u+p_1}-\frac{1}{u+p_3}\right)
\right] +
$$
$$
+
q_2
\left[
e^{\bar{u}z - u\bar{z}} \bar{u}
\left(\frac{1}{\bar{u}-p_2} - \frac{1}{\bar{u}-p_3}\right)
-
e^{\bar{u}\bar{z}-uz} u
\left(\frac{1}{u+p_2}-\frac{1}{u+p_3}\right)
\right]  =
$$
$$
=
e^{\bar{u}\bar{z}-uz}\frac{u}{u+p_3}
-
e^{\bar{u}z-u\bar{z}}\frac{\bar{u}}{\bar{u}-p_3}.
$$
The formula for the potential $V=\xi^-_1$ takes the form
$$
V = |u|^2\left(\frac{q_1}{p_1}+\frac{q_2}{p_2}+\frac{1-q_1-q_2}{p_3}\right).
$$

We can also compute $t_1$ and $t_2$ solving analogous equations.
The formula for the potential $U$ is
\begin{equation}
\label{potential}
U = t_1 p_1 + t_2 p_2 + (1-t_1-t_2) p_3.
\end{equation}

In both cases we have to know the points $p_1,p_2,p_3$ to make computations.
It follows from Theorem \ref{theorem2} that if these points are chosen 
in a rather convenient way then the potentials $U$ and $V$ are real-valued
and moreover coincide $U=V$.
This is the case when 
there are meromorphic differentials $\omega$ and $\omega^\prime$
with properties exposed in Theorem \ref{theorem2}.

Let us look for such differentials.

The general form of a differential $\omega$ on $\Gamma$ which has poles at
$\infty_\pm$ with the principal parts
$\pm k_\pm^2 (1+O(k^{-1}_\pm))dk^{-1}_\pm$
and is regular outside these points is
$$
\omega =
-
\left[
\frac{\lambda^2-|u|^2}{\lambda^2} + a
\left(
\frac{1}{\lambda-u}-\frac{1}{\lambda+\bar{u}}
\right)
+ b
\left(
\frac{1}{\lambda+u}-\frac{1}{\lambda-\bar{u}}
\right)
\right]
d\lambda
$$
and the general form of
a differential $\omega^\prime$ on $\Gamma$ which has poles at $\infty_\pm$ with
the principal parts
$k_\pm^2 (1+O(k^{-1}_\pm))dk^{-1}_\pm$
and is regular outside these points is
$$
\omega^\prime =
-
\left[
\frac{\lambda^2+|u|^2}{\lambda^2}
+ c \left(
\frac{1}{\lambda-u}-\frac{1}{\lambda+\bar{u}}
\right)
+ d
\left(\frac{1}{\lambda+u}-\frac{1}{\lambda-\bar{u}}
\right)
\right]
d\lambda.
$$
We have to find differentials $\omega$ and $\omega^\prime$ such that their
zero sets would be of the form $D+\sigma(D)$ and $D+\tau(D)$ respectively.

Notice that the zero sets of $\omega$ and $\omega^\prime$ are the zero sets of
the polynomials $Q(\lambda)$ and $Q^\prime(\lambda)$ respectively where
$$
Q(\lambda) = (\lambda^2-|u|^2)(\lambda^2-u^2)(\lambda^2-\bar{u}^2) +
$$
$$
+
\lambda^2(u+\bar{u})[(a-b)(\lambda^2-|u|^2) +(a+b)(u-\bar{u})\lambda],
$$
$$
Q^\prime(\lambda) = (\lambda^2+|u|^2)(\lambda^2-u^2)(\lambda^2-\bar{u}^2) +
$$
$$
+
\lambda^2(u+\bar{u})[(c-d)(\lambda^2-|u|^2) +(c+d)(u-\bar{u})\lambda].
$$
Since we look for a differential $\omega$ with
the zero set of the form $D+\sigma(D)$
the polynomial $Q(\lambda)$ has to have only terms with even powers.
This implies that $b=-a$ and we have
$$
Q(\lambda) = (\lambda^2-|u|^2)[(\lambda^2-u^2)(\lambda^2-\bar{u}^2) +
2a(u+\bar{u})\lambda^2].
$$
We see that $Q(|u|)=0$ and put
$$
p_3 = |u|.
$$
Other poles $p_1$ and $p_2$ satisfy the equation
$$
(\lambda^2-u^2)(\lambda^2-\bar{u}^2) +
2a(u+\bar{u})\lambda^2 = 0
$$
together with $-p_1,-p_2$.
Since $\tau(p_3)=p_3$, the point $p_3=|u|$ has to be at least
a double root of the polynomial $Q^\prime(\lambda)$.

We skip some simple calculations which show that if we assume that $c=d$
and that $(\lambda-|u|)^2$ divides $Q^\prime(\lambda)$ then we have
$$
Q^\prime(\lambda) = (\lambda-|u|)^2\left[\lambda^4 + 2|u|\lambda^3 +
(4|u|^2-(u^2+\bar{u}^2))\lambda^2 + 2|u|^3\lambda + |u|^4 \right] =
$$
and
$$
c=d=\frac{(u^2+\bar{u}^2)|u|-2|u|^3}{u^2-\bar{u}^2}.
$$
We have
$$
Q^\prime(\lambda) = (\lambda-|u|)^2 |u|^4 P\left(\frac{\lambda}{|u|}\right),
$$
where
$$
P(\mu)=(\mu^2-A\mu+1)(\mu^2-\bar{A}\mu+1),
$$
$$
A+\bar{A} = -2, \ \ |A|^2 = 2 - \frac{u^2+\bar{u}^2}{|u|^2}.
$$
and
$$
A = t + t^{-1}, \ \ \bar{A} = \bar{t}+\bar{t}^{-1},
$$
where $t,\bar{t},t^{-1}$ and $\bar{t}^{-1}$ are the roots of the polynomial
$P(\mu)$.

Until now we did all computations for a general value of $u$. let us
consider the special case when
$$
u = \frac{1+i}{4}, \ \ \ u^2+\bar{u}^2 = 0.
$$
The polynomial $P(\mu)$ takes a very simple form and we can easily 
find all its roots:
$$
P(\mu) = \mu^4 + 2\mu^3 + 4\mu^2 + 2\mu +1,
$$
$$
\mu_{1,2} = \frac{-(1-i) \pm \sqrt{-2i-4}}{2}, \ \ \ 
\mu_{3,4}= \bar{\mu}_{1,2} = \frac{-(1+i) \pm \sqrt{-2i-4}}{2}
$$
and we obtain four other zeros of the differential $\omega^\prime$:
$$
s_1 = \frac{-1+i + \sqrt{-2i-4}}{4\sqrt{2}}, \ \ \
s_2 = \frac{-1+i - \sqrt{-2i-4}}{4\sqrt{2}}, 
$$
$$
\tau(s_1) = \frac{-1-i-\sqrt{2i-4}}{4\sqrt{2}}, \ \ \
\tau(s_2) = \frac{-1-i-\sqrt{2i-4}}{4\sqrt{2}}.
$$
Therefore for $c=d=\frac{i}{\sqrt{8}}$ the differential $\omega^\prime$ has 
zeroes exactly at $D+\tau(D)$ where $D=s_1+s_2+p_3$ where $p_3=|u|$ and
$\tau(p_3)=p_3$.

Notice that the zeros of $\omega$ have to be invariant with respect to 
the involution $\sigma$ and their product equals $|u|^6$.
This implies that we can put either $p_1=s_1, p_2=s_2$ either
$p_1=\tau(s_1), p_2=\tau(s_2)$. 

Let us make our choice as follows:
$$
p_1 = \frac{-1+i + \sqrt{-2i-4}}{4\sqrt{2}}, \ \
p_2 = \frac{-1+i - \sqrt{-2i-4}}{4\sqrt{2}}, \ \
p_3 = |u| = \frac{1}{\sqrt{8}}.
$$
In this event $a=-b=\frac{1+i}{\sqrt{8}}$.

Substituting these values of $p_k,k=1,2,3$, into the equations for $q_1$ and 
$q_2$ we compute them as functions of $x,y$ and by that obtain the explicit 
formula for $\psi_1$ and $V$. Now the functions $\psi_2$ and $U$ can be found
by solving analogous equations for $t_1,t_2$ or from the following formulas
which follows from Theorem \ref{theorem2}:
$$
U=V, \ \ \ \psi_2 = \frac{\bar{\partial} \psi_1}{V}.
$$
We omit the explicit formulas for $q_1,q_2,t_1,t_2$ and just mention 
that these functions appear to be
functions of $y$ which are $2\pi$-periodic. The $\psi$-function at the point
$\lambda=u$ equals:
$$
\psi_1(z,\bar{z},\bar{u}) = \frac{1+i}{4} e^{-\frac{ix}{2}} 
\frac{2 e^{\frac{iy}{2}} + \sqrt{2}(1-i) e^{-\frac{iy}{2}}}
{\sin y - \sqrt{2}},
$$
$$ 
\psi_2(z,\bar{z},\bar{u}) = -\frac{\sqrt{2}}{4}e^{-\frac{ix}{2}}
\frac{2 e^{\frac{iy}{2}} 
-\sqrt{2}(1+i)e^{-\frac{iy}{2}}}
{\sin y - \sqrt{2}}.
$$
We see that
$$
i\bar{\psi}_2^2 = 
- \frac{(1+i)e^{ix}}{2}\frac{\sqrt{2} -\cos y -\sin y}{(\sin y-\sqrt{2})^2},
$$
$$
\psi_1 \bar{\psi}_2 = 
\frac{\sqrt{2}i}{2} \frac{1-\sqrt{2} \sin y}{(\sin y -\sqrt{2})^2}
$$
and comparing these formulas with the formulas (\ref{clifform2})
we conclude that a surface constructed from the function
$\frac{1}{\root 4 \of{2}} \psi(z,\bar{z},\bar{u})$ by the Weierstrass
representation is a torus which is mapped
to the Clifford torus defined by the formulas (\ref{clifform1}) 
by the orthogonal transformation
$$
T = 
\left(
\begin{array}{ccc}
- \frac{1}{\sqrt{2}} & \frac{1}{\sqrt{2}} & 0 \\
-\frac{1}{\sqrt{2}} & -\frac{1}{\sqrt{2}} & 0 \\
0 & 0 & -1
\end{array}
\right).
$$
This transformation changes the orientation of $\R^3$ which 
results in the change of the sign of the potential $U$: the formula
(\ref{potential}) gives us
$$
U = \frac{\sin y}{2\sqrt{2}(\sin y - \sqrt{2})}
$$
and this expression differs by the sign from (\ref{clifpotential}).

\end{document}